\documentclass[preprint,12pt]{elsarticle}

\usepackage{amssymb}
\usepackage{graphicx}
\usepackage{booktabs}
\usepackage{tabularx}
\usepackage{array}
\usepackage{lscape}
\usepackage{longtable}
\usepackage{multirow}
\usepackage{amsmath}
\usepackage{colortbl}
\usepackage{subfigure}
\usepackage{url}

\newcommand{\PreserveBackslash}[1]{\let\temp=\\#1\let\\=\temp}
\newcolumntype{C}[1]{>{\PreserveBackslash\centering}p{#1}}
\newcolumntype{R}[1]{>{\PreserveBackslash\raggedleft}p{#1}}
\newcolumntype{L}[1]{>{\PreserveBackslash\raggedright}p{#1}}

\journal{Physics letter A} 
\begin{document}

\begin{frontmatter}



\title{A new information dimension of complex networks}


\author[address1,address2]{Daijun Wei}
\author[address1]{Bo Wei}
\author[address3]{Yong Hu}
\author[address1]{Haixin Zhang}
\author[address1,address4]{Yong Deng \corref{label1}}
\address[address1]{School of Computer and Information Science, Southwest University, Chongqing 400715, China}
\address[address2]{School of Science, Hubei University for Nationalities, Enshi 445000, China}
\address[address3]{Institute of Business Intelligence and Knowledge Discovery, Guangdong University of Foreign Studies, Guangzhou 510006, China}
\address[address4]{School of Engineering, Vanderbilt University, TN 37235, USA}
\cortext[label1]{Corresponding author: School of Computer and Information Science, Southwest University, Chongqing 400715, China.  Tel:+86 023-68254555; E-mail address: ydeng@swu.edu.cn;prof.deng@hotmail.com}

\begin{abstract}
The fractal and self-similarity properties are revealed in many real complex networks. However, the classical information dimension of complex networks is not practical for real complex networks. In this paper, a new information dimension to characterize the dimension of complex networks is proposed. The difference of information for each box in the box-covering algorithm of complex networks is considered by this measure. The proposed method is applied to calculate the fractal dimensions of some real networks. Our results show that the proposed method is efficient for fractal dimension of complex networks.
\end{abstract}


\end{frontmatter}


\section{Introduction}
Recently, complex networks have attracted a growing interest in many disciplines \cite{newman2003structure,yu2009pinning,zheng2009adaptive,song2010synchronization,vidal2011interactome,wei2013networks,zhang2013impulsive}.
Several  properties of complex networks have been
revealed, including small-world phenomena \cite{watts1998collective}, scale-free
degree \cite{barabasi1999emergence} and community structure \cite{fortunato2010community} etc. The fractal theory has been used in the study of various subjects \cite{mandelbrot1984fractal,suzuki1990fractal,doering2006multiscale,calcagni2010fractal,imakaev2012fractal}. The fractal and self-similarity of complex networks are discovered by Song et al.\cite{song2005self}. From then on, the fractal and self-similarity of complex networks are extensively studied by many researchers \cite{long2009fractal,lu2009similarity,vitiello2012fractals,comellas2013number,hwang2013origin,silva2012local,wu2013controlling}. The box-covering algorithm is described in detail and applied to demonstrate the existence of self-similarity in many real complex networks \cite{song2006origins,song2007calculate}. From then on, the box-covering algorithm for the complex networks is extensively studied. For instance, the traditional box-covering algorithm is improved in the references  \cite{kim2007fractality,gallos2007review,gao2008accuracy,ng2011growth,schneider2012box,hxzhang2013selfsimilarity} and modified for weighted complex networks \cite{wei2013box}.

The fractal and self-similarity properties of complex are revealed from other perspectives, such as volume dimension \cite{shanker2007defining,shanker2007graph}, correlation dimension \cite{lacasa2013correlation} and information dimension \cite{zhuhua2011fractals}. However, In the classical information dimension of complex networks, complex networks must be measured in plane area \cite{zhuhua2011fractals}. It is almost impossible for real complex networks since the distance between nodes can not be obtained in the plane.  We find that most of boxes cover different number of nodes for given a box size in the box-covering algorithm. It means that different information are contained even in the boxes of same size. In this paper, combining with the classical information dimension and the box-covering algorithm, a new information dimension to characterize the dimension of complex networks is proposed.
In what follows,  the classical information dimension of complex networks is introduced in section 2. The proposed method of information dimension for complex networks is depicted in section 3. In section 4, the efficiency of the proposed method is illustrated by calculating fractal
dimension of some real complex networks. Some conclusions are presented in section 5.

\section{The classical information dimension of complex networks}
In this section, the classical information dimension of complex networks is briefly introduced. Information dimension was introduced by Renyi based on the probability method \cite{renyi1960dimension}. The ideal of the classical information dimension is that the fractal target is covered by varies sizes of squares. For a give square size $\varepsilon$, the information equation is given as follows \cite{zhuhua2011fractals},
\begin{equation}\label{inf equation}
I =- \sum\limits_{i = 1}^{{N_{\varepsilon}}} {{p_i}(\varepsilon)\ln{p_i}(\varepsilon)}
\end{equation}
where ${p_i}(\varepsilon)$ is the probability of the fractal target contained in the ith box, $N_{\varepsilon}$ is number of all boxes. The information dimension is defined as follows \cite{zhuhua2011fractals},
\begin{equation}
{d_i} = -\mathop {\lim }\limits_{\varepsilon \to 0} \frac{{I}}{{\ln{\varepsilon}}}=\mathop {\lim }\limits_{\varepsilon \to 0} \frac{{\sum\limits_{i = 1}^{{N_{\varepsilon}}} {{p_i}(\varepsilon)\ln{p_i}(\varepsilon)}}}{{\ln{\varepsilon}}}
\end{equation}
where $d_i$ is the information dimension of the fractal target. For complex network $G={(N, V)}$ and $N={(1,2,\cdots,n)}$,
$V={(1,2,\cdots,m)}$, where $n$ is the total number of nodes and $m$ is the total number of edges. $d_i$ is the information dimension of complex network when a complex network is covered by varies size of squares in the plane area. For given square size $\varepsilon$, $p_{i}(\varepsilon)$ is given as \cite{zhuhua2011fractals}
\begin{equation}
{p_i}(\varepsilon ) = \frac{{{q_i}(\varepsilon )}}{{{Q_i}(\varepsilon )}}
\end{equation}
where ${q_i}(\varepsilon )$ is number of squares including nodes of complex network, ${Q_i}(\varepsilon )$ is the number of squares covering the given plane area and ${p_i}(\varepsilon )$ is ration of number of squares. The given plane area is always fixed. Numbers of ${q_i}(\varepsilon )$ and ${Q_i}(\varepsilon )$ are varied with squares size $\varepsilon$.
In theory, the classical information dimension of complex network is feasible. However, it is unreasonable that a real complex network is embedded in the plane area since the distance between nodes can not be obtained. It is almost impossible to obtain the value of ${p_i}(\varepsilon )$.

\section{A new information dimension of complex networks}
\subsection{The box-covering algorithm of complex networks}
\begin{figure}[!ht]
\begin{center}
\includegraphics[width=9cm,height=6cm]{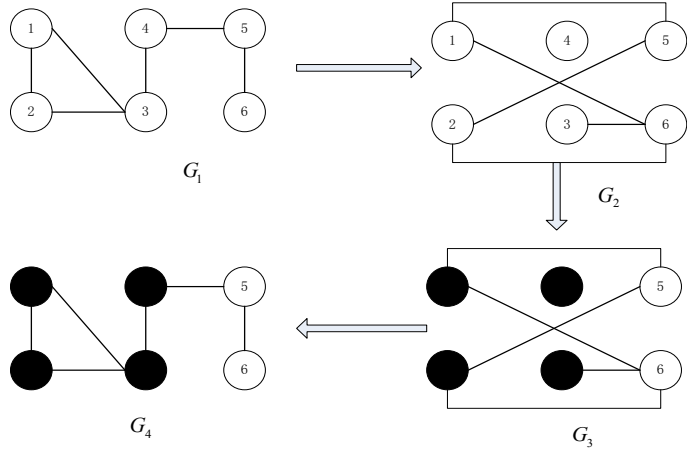}
\caption{The idea of the classical box-covering algorithm for complex networks, where $l=3$}
\label{fp1}       
\end{center}
\end{figure}

The original definition of box-covering is initially proposed by Hausdorff
\cite{jurgens1992chaos,bunde1994fractals}. It is  applied
in complex networks by Song, et al. \cite{song2005self,song2007calculate}.  The idea of the box-covering algorithm of complex networks has
three steps showing in Figure \ref{fp1}
\cite{song2007calculate}.

Step 1: For a given network $G_{1}$ and a box size $l$, a new
network $G_{2}$ is obtained, in which node $i$ is connected to
node $j$ when $d_{ij} \ge l$.

Step 2: Mark each node with a free color, which is different from
the colors its nearest neighbours in the network $G_{2}$. And then, a
network $G_{3}$ is obtained.

Step 3: Let one color represent one box in the network $G_{4}$. And
then, the value of box ($N_{b}(l)$) is obtained.

For a given box size $l$, randomly assign a unique id from 1 to $n$ to all network nodes, every box is a set of nodes where all
distances $d_{ij}$ between any two nodes $i$ and $j$ in the box
are smaller than $l$. The minimum number of boxes $N_{b}(l)$ must cover the entire network.  Increase $l$ by one until $l$ is more than $max(d_{ij})$. For fractal complex networks, the relationship $N_{b}(l)$ and $l$ is given as follows,
\begin{equation}\label{fractal}
{N_b}(l) \sim {l}^{-{d_b}}
\end{equation}
where $d_{b}$ is box dimension of this network. The value of $d_{b}$ is obtained as follows \cite{song2007calculate},
\begin{equation}
{d_{b}}= -\mathop {\lim }\limits_{l \to 0}  \frac{{\ln {N_b}(l)}}{{\ln {l}}}.
\end{equation}
As described above, the distance between nodes only depend on number of edges, which connect from a node to another. Complex networks do not embed the plant area. The algorithm has been widely used to calculate fractal dimension of complex networks.
%
%
%
%
%
%

\subsection{Definition of proposed information dimension for complex networks}
 In our proposed information dimension of complex networks, the classical information dimension and the box-covering algorithm are referenced. Most of boxes have different number of nodes for given a box size in the box-covering algorithm. The difference number of nodes is considered in our method. For a give box size $l$, the probability of information in the ith box is denoted as $p_{i}^{'}(l)$ and defined as follows,
\begin{equation}\label{inf}
{p_i}^{'}(l) = \frac{{{n_i}(l)}}{n}
\end{equation}\label{inf2}
where $n_{i}(l)$ is the number of nodes in the ith box and $n$ is number of complex networks. Similar to Equation \ref{inf equation}, information equation is given as follows,
\begin{equation}
I^{'}(l) =- \sum\limits_{i = 1}^{{N_b}} {{p_i}^{'}(l)\ln {p_i}^{'}(l)}.
\end{equation}
A complex network has fractal property based on information dimension if it satisfies
\begin{equation}
I^{'}(l) \sim {l^{-{d_i}^{'}}}
\end{equation}
where ${d_i}^{'}$ is information dimension of complex networks. ${d_i}^{'}$ is obtained as follows,
\begin{equation}\label{inf1}
{{d_i}^{'}} = -\mathop {\lim }\limits_{l \to 0} \frac{{I^{'}(l)}}{{\ln (l)}}=\mathop {\lim }\limits_{l \to 0} \frac{{\sum\limits_{i = 1}^{{N_b}} {{p_i}^{'}(l)\ln {p_i}^{'}(l)}}}{{\ln (l)}}.
\end{equation}
Using Equations \ref{inf} and \ref{inf1}, we have
\begin{equation}\label{inf3}
{d_i}^{'} = \mathop {\lim }\limits_{l \to 0} \frac{{\sum\limits_{i = 1}^{{N_b}} {\frac{{{n_i}(l)}}{n}\ln \frac{{{n_i}(l)}}{n}} }}{{\ln (l)}}
\end{equation}

In this definition of information dimension of complex networks, the probability of information is represented the ration of node number. The number of nodes is easily measured by the box-covering algorithm. The information dimension of real complex networks is achievable based on the box-covering algorithm.
\section{Applications and discussions}
\begin{figure}[!t]
\centering
\subfigure[The email network] {\includegraphics[width=2.3in]{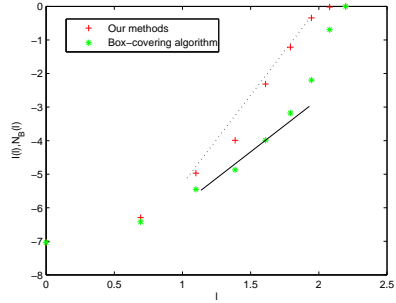}}
\subfigure[The American college football] {\includegraphics[width=2.3in]{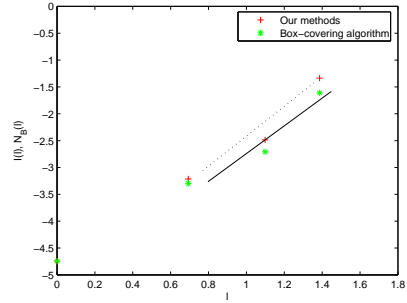}}
\subfigure[The dolphins social network] {\includegraphics[width=2.3in]{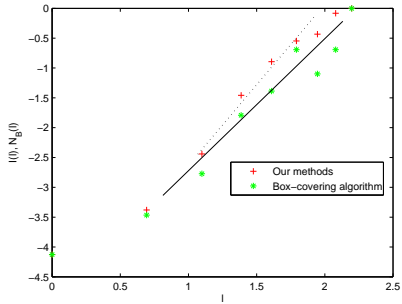}}
\subfigure[The power network] {\includegraphics[width=2.3in]{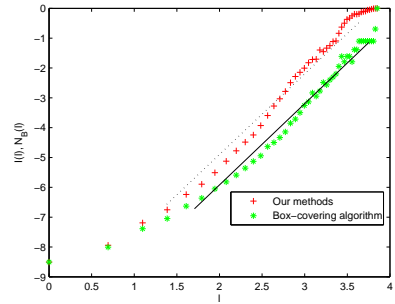}}
\caption{ Fractal scaling analysis of these real complex networks. Symbols refer to: * and + indicate the correlation between $l$ and $N_{b}$ and $l$ and $I^{'}(l)$ by using box-covering algorithm and our method, respectively. }
\label{fig}
\end{figure}

In the section, the email network (http://vlado.fmf.unilj.si/pub/networks/data/), the American college football, the dolphins social network and the power network (http://www-personal.umich.edu/mejn/netdata/) are calculated by using our method and the box-covering algorithm \cite{song2007calculate}. The number of nodes and edges are given in the first and second row of Table \ref{shuju}. The fractal dimension is averaged for 1000 times. Fractal scaling analysis of these real complex networks is shown in Figure \ref{fig}.
By means of the least square fit, the reference lines have different slopes, respectively.  $Q$ represents sum square error (SSE) of linear \cite{pedrycz2003fuzzy}. $Q$ is defined as follows,
\begin{equation}
Q = {\sum\limits_k {({y_k} - \mathop {{y_k}}\limits^ \wedge  )} ^2}
\end{equation}
where $y_k$ are values of discrete points and $\mathop {{y_k}}\limits^ \wedge$ is a function of line. $Q_i$ and $Q_b$ represent SSE of linear by our method and Song's method \cite{song2007calculate}, respectively.
The values of ${d_i}^{'}$ , $Q_i$, $d_b$ and $Q_b$ for these real networks are shown in Table \ref{shuju}.
\begin{table}[htbp!]
\begin{center}
\caption{General characteristics of sever real networks and the fractal dimension ${d_i}^{'}$ by our proposed method and $d_b$ by Song's method \cite{song2007calculate}.  $Q_i$ and $Q_b$ represent SSE of linear in our method and Song's method \cite{song2007calculate}, respectively.}
\label{shuju}
\begin{tabular}{|l|c|c|c|c|c|c|c|} \hline
Network & Size &edges & ${d_i}^{'}$ & $Q_i$&$d_b$ &$Q_b$\\
 \hline
Email network & 1133 &10902& 4.838& 0.751& 3.833& 1.995 \\\hline
American college football & 115 & 615 & 2.766 &0.0846& 2.688& 1.0442\\\hline
Dolphins social network& 62 & 159 & 2.061&0.6018& 1.888& 0.7229\\\hline
Power grid & 4941&6594 & 2.694&2.4877& 2.4109&3.0832 \\\hline
\end{tabular}
\end{center}
\end{table}

From Figure \ref{fp1} and Table \ref{shuju}, the email network, the American college football and the dolphins social network have significant  fractal property. However, the fractal property of the power gird is not apparent. This conclusion about the power gird is consistent with Ref \cite{kim2007fractality}. From Table \ref{shuju}, the values of SSE of our method are small. It demonstrates that the information dimension is effective for fractal dimension of complex networks. In general, we can see that both our proposed method and Song's method could capture the fractal property of these real complex networks.
\section {Conclusions}
It is an important issue that how to describe the fractal and self-similarity of complex network. Some different physical quantities are considered in the definitions of fractal, such as the box-covering fractal dimension \cite{song2007calculate} and the classical information dimension \cite{zhuhua2011fractals}. The box-covering algorithm has been used widely for fractal dimension of complex networks. However, the classical information dimension of complex networks is calculated rarely because of the lack of operability. In this paper, a new and easy-operating algorithm of information dimension is proposed based on the box-covering algorithm for complex networks. Computational results of real networks show that the proposed method is effective and flexible. The proposed method is useful to reveal fractal property of complex networks.

\section*{Acknowledgment}
 The work is partially supported Chongqing Natural Science Foundation, Grant No. CSCT, 2010BA2003, National Natural Science Foundation of China, Grant Nos. 61174022 and 61364030, National High Technology Research and Development Program of China (863 Program) (No.2013AA013801), the Fundamental Research Funds for the Central Universities, Grant No. XDJK2012D009.

\bibliographystyle{model1-num-names}
\bibliography{weighted}


\end{document}